\documentstyle[twoside,fleqn]{article}

\def\soc{{\rm C}_{60}}
\def\rug{{\rm C}_{70}}
\def\eps{\varepsilon}

\title{Optical response of C$_{60}$ and C$_{70}$ fullerenes:
Exciton and lattice fluctuation effects}

\author{Kikuo Harigaya and Shuji Abe\\ \\
       Fundamental Physics Section, Electrotechnical Laboratory,\\
       Umezono, Tsukuba, Ibaraki 305, Japan
}

\begin{document}

\begin{abstract}
Molecular exciton effects in the neutral and maximally doped C$_{60}$
(C$_{70}$) are considered using a tight binding model with long-range
Coulomb interactions and bond disorder.  By comparing calculated and
observed optical spectra, we conclude that relevant Coulomb parameters
for the doped cases are about half of those of the neutral systems.
The broadening of absorption peaks is well simulated by the bond disorder
model.
\end{abstract}


\maketitle

\section{INTRODUCTION}
\small

Fullerenes have a hollow cage of carbons \cite{1}.  Its structure
is formed with $\sigma$-bondings.  The $\pi$-electron orbitals
extend to the outside of the cage.  As $\pi$-electrons are delocalized
on the surface of fullerenes, the optical properties are similar to
those of $\pi$-conjugated polymers \cite{2}.  Recently, we have applied
the intermediate exciton formalism which was used in polymers \cite{3},
and have investigated relevant Coulomb parameters in order to understand
the dispersions and oscillator strengths of the absorption spectra
of the neutral $\soc$ ($\rug$) \cite{4,5} and maximally doped
systems \cite{6,7}.  The purpose of this article is to review the
main results published elsewhere \cite{8,9}.

In the formalism \cite{8}, a tight binding model with a long-range Coulomb
interaction and bond disorder has been used.  The free electron
part corresponds to the simple H\"{u}ckel model with the mean hopping
integral $t$; the presence of the dimerization is considered for
the neutral $\soc$ case, and the constant hopping is assumed for the
other cases.  The long-range Coulomb interaction is the Ohno
expression: $W(r) = 1/\sqrt{(1/U)^2 + (r/r_0 V)^2}$.  Here, $U$
is the onsite Coulomb strength, $V$ is the long-range component,
and $r_0$ is the average bond length.  The bond disorder model
with the Gaussian distribution of the hopping integral with a
standard deviation $t_{\rm s}$ simulates the lattice fluctuation
effects \cite{10}.  This gives rise to broadening of the absorption
spectra.  The model is solved by the Hartree-Fock approximation and
the single excitation configuration interaction method.  Optical
spectra are obtained by using the dipole approximation.  In the
next two sections, we present the calculated spectra and compare them
with experiments.

\section{NEUTRAL C$_{\bf 60}$ AND C$_{\bf 70}$}

We discuss the calculated optical absorption spectra of $\soc$ and $\rug$.
$\soc$ has high $I_h$ symmetry, and $\rug$ has lower $D_{5h}$ symmetry.
Figure 1(a) shows the spectrum of $\soc$, and Fig. 1(b) that of $\rug$.
Parameters, $U=4t$, $V=2t$, $t_{\rm s} = 0.09t$, and $t=1.8$eV, are used.
These parameters are determined so as to reproduce the experimental spectra
as good as possible.  Experimental data of molecules in solutions
are taken from \cite{4,5}, and are shown by thin lines.
There are three main features in $\soc$ absorption found around
the energies 3.5eV, 4.7eV, and 5.6eV.  In the case of $\rug$,
we could say that several small peaks in the energy interval
from 1.7eV to 3.6eV are originated from the 3.5eV feature of $\soc$ after
splitting.  The 4.7eV and 5.6eV features of $\soc$ overlap
into the large feature present in the energy region
larger than 3.6eV.  The optical gap decreases from 3.1eV
($\soc$) to 1.7eV ($\rug$).  These changes would be due to the
symmetry reduction from the $\soc$ soccer ball to the $\rug$ rugby ball.
There is a good agreement with experiments about the peak positions
and relative oscillator strengths in the $\soc$ case.  The agreement
is overall for $\rug$.  There is a systematic deviation of the
experimental curves from the theoretical ones at energies higher
than 5.0eV due to the overlap of $\sigma$-electron excitations.

\begin{figure}[htb]
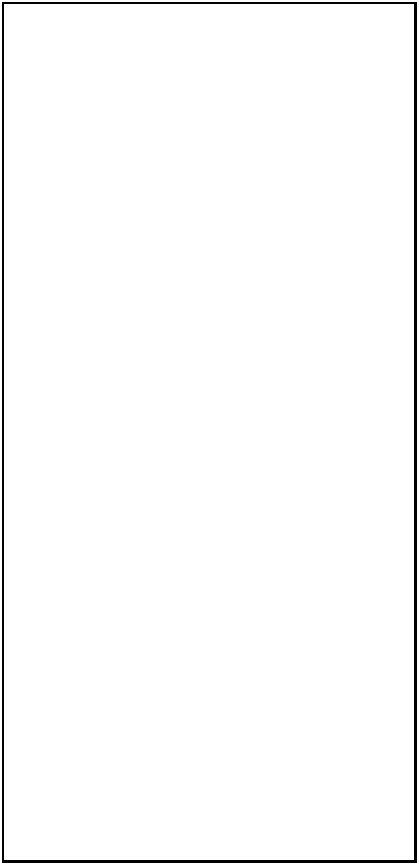

\vspace{9pt}
\framebox[55mm]{\rule[-21mm]{0mm}{112mm}}
\caption{Optical absorption spectra for (a) C$_{60}$ and (b) C$_{70}$,
shown in arbitrary units.  The parameters in the Ohno potential
are $U = 4t$ and $V = 2t$ ($t=1.8$eV).   Experimental data are shown
by thin lines.  They are taken from \protect\cite{4} for (a),
and from \protect\cite{5} for (b).}
\label{fig:1}
\end{figure}

\section{MAXIMALLY DOPED C$_{\bf 60}$ AND C$_{\bf 70}$}

Calculations are performed for $\soc^{6-}$ and $\rug^{6-}$ in order to
look at molecular exciton effects in $A_6 \soc$ and $A_6 \rug$ solids ($A =$
alkali metals).  Results are shown in Figs. 2(a) and (b) with the experimental
absorption \cite{6} and the optical conductivity data obtained by the
EELS studies \cite{7}.  We find that the parameter set, $U=2t$, $V=1t$,
$t_{\rm s} = 0.20t$, and $t=2.0$eV, is relevant to both systems.
For $A_6 \soc$, two features around 1.2eV and 2.8eV are reasonably
explained by the present calculation.  The broadening is simulated well
by the bond disorder.  The disorder strength $t_{\rm s} = 0.20t$ is
about the twice as large as that of the neutral $\soc$.  The broad feature
in the energies higher than 3.6eV in the experiments might correspond to the
two broad peaks around 4.4eV and 6.0eV of the calculation.  In these
energies, the agreement is not so good.  The same fact has been seen
in the neutral $\soc$ case.  Excitations which include $\sigma$-orbitals
would mix in this energy region.  This effect could be
taken into account by using models with $\pi$- and $\sigma$-electrons.
For $A_6 \rug$ shown in Fig. 2(b), we can assign that the features
at 1.2eV, 2.6eV, and 4.6eV of the calculation correspond to
the peaks at 1.0eV, 2.7eV, and 4.2eV of the experiment, respectively.
Thus, we can explain the presence of three broad features observed by
experiments.

\begin{figure}[htb]
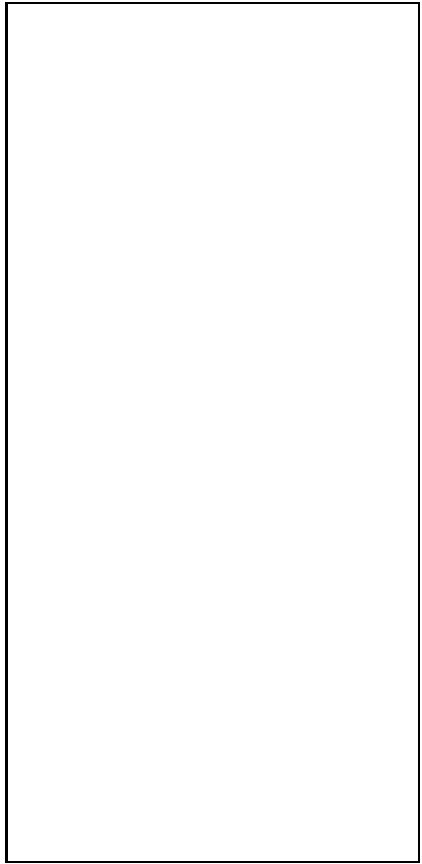

\vspace{9pt}
\framebox[55mm]{\rule[-21mm]{0mm}{112mm}}
\caption{Optical absorption spectra for
(a) C$_{60}^{6-}$ and (b) C$_{70}^{6-}$,
shown in arbitrary units.  The parameters in the Ohno potential
are $U = 2t$ and $V = 1t$ ($t=2.0$eV).   Experimental data of maximally
doped systems are taken from \protect\cite{6} (thin line) and
\protect\cite{7} (dots)  for (a), and from \protect\cite{7} (dots) for (b).}
\label{fig:2}
\end{figure}

\section{DISCUSSION}

We have mainly looked at Frenkel exciton effects in the neutral and
maximally doped $\soc$ (and $\rug$) molecules.  The relevant Coulomb
parameters for the optical spectra are $U \sim 2V \sim 4t$ for the
neutral systems, and $U \sim 2V \sim 2t$ for $A_6 \soc$ and $A_6 \rug$
solids.  In the Ohno potential, the dielectric screening of Coulomb
interactions can be taken into account by assuming $U = U_0/\eps_{\rm s}$
and $V = V_0/\eps_{\rm l}$, where $\eps_{\rm s}$ and $\eps_{\rm l}$
are short- and long-range dielectric constants.  The magnitude of the
static dielectric constant increases by the factor about two upon doping:
$\eps_1(0) = 4.3$ in the neutral $\soc$ and $\eps_1(0) = 7.1$
in Rb$_6\soc$, and also $\eps_1(0) = 4.0$
in the neutral $\rug$ and $\eps_1(0) = 8.0$ in Rb$_6\rug$ \cite{7}.
This enhancement of dielectric constants reasonably accords with
the decrease of $U$ and $V$ in the present calculations.


\begin{thebibliography}{99}
\bibitem{1} H.W. Kroto, J.E. Fischer and D.E. Cox (eds.),
The Fullerenes, Pergamon Press, Oxford, 1993.
\bibitem{2} A.J. Heeger, S. Kivelson, J.R. Schrieffer and
W.P. Su, Rev. Mod. Phys. 60 (1988) 781.
\bibitem{3} S. Abe, J. Yu and W.P. Su, Phys. Rev. B
45 (1992) 8264.
\bibitem{4} S.L. Ren, Y. Wang, A.M. Rao, E. McRae, J.M. Holden,
T. Hager, K.A. Wang, W.T. Lee, H.F. Ni, J. Selegue and P.C. Eklund,
Appl. Phys. Lett. 59 (1991) 2678.
\bibitem{5} J.P. Hare, H.W. Kroto and R. Taylor, Chem. Phys. Lett.
177 (1991) 394.
\bibitem{6} T. Pichler, M. Matus, J. K\"{u}rti and H. Kuzmany,
Solid State Commun. 81 (1992) 859.
\bibitem{7} E. Sohmen and J. Fink, Phys. Rev. B 47 (1993) 14532.
\bibitem{8} K. Harigaya and S. Abe, Phys. Rev. B 49 (1994) 16746.
\bibitem{9} K. Harigaya, Jpn. J. Appl. Phys. 33 (1994) in press.
\bibitem{10} K. Harigaya, Phys. Rev. B 48 (1993) 2765.


\end{thebibliography}
\end{document}